# Signal-to-noise and spatial resolution in in-line imaging. 1. Basic theory, numerical simulations and planar experimental images

T. E. Gureyev,[1,2,*] , D. M. Paganin[2] and H. M. Quiney[1]

[1]*School of Physics, the University of Melbourne, Parkville, VIC 3010, Australia*
[2]*School of Physics and Astronomy, Monash University, Clayton, VIC 3800, Australia*
*\*timur.gureyev@unimelb.edu.au*

**Abstract**

Signal-to-noise ratio and spatial resolution are quantitatively analysed in the context of in-line (propagation based) X-ray phase-contrast imaging. It is known that free-space propagation of a coherent X-ray beam from the imaged object to the detector plane, followed by phase retrieval in accordance with Paganin's method, can increase the signal-to-noise in the resultant images without deteriorating the spatial resolution. This results in violation of the noise-resolution uncertainty principle and demonstrates "unreasonable" effectiveness of the method. On the other hand, when the process of free-space propagation is performed in software, using the detected intensity distribution in the object plane, it cannot reproduce the same effectiveness, due to the amplification, during free-space propagation, of photon shot noise in the object-plane intensity. We show that the performance of Paganin's method is determined by just two dimensionless parameters: the Fresnel number and the ratio of the phase shift to the logarithm of intensity in the object plane. The relevant theoretical analysis is performed first, followed by computer simulations and then by a brief test using experimental images collected at a synchrotron beamline. More extensive experimental tests will be presented in the second part of this paper.

## 1. Introduction

X-ray phase-contrast imaging (PCI) is a class of powerful techniques for analysing the internal structure of non-crystalline samples [1-4]. Seminal results in this field were obtained in the 1990s, using coherent beams generated by synchrotrons [5,6] and laboratory sources [7-9], although some closely related experiments were demonstrated earlier [10-12]. X-ray PCI has a significant potential for increasing image contrast of biological soft tissues, compared to conventional absorption-based X-ray imaging. Various forms of X-ray PCI of biological samples have been demonstrated over the years [2-4]. In this work, however, we focus on just one method, known as in-line or propagation-based imaging (PBI) [5,9,13,14]. For a detailed history of development of X-ray PCI technology and the current state of the art see, for example, Refs. [1-4].

PBI represents the simplest X-ray phase-contrast imaging technique, at least in principle. It does not require any optical elements in order to render phase contrast visible, relying instead on the free-space propagation of the beam transmitted through the sample before it is registered by a position-sensitive detector. The key requirement for PBI is a sufficiently high degree of spatial coherence of the illuminating beam [1,8,14], which is typically achieved by using either highly-collimated synchrotron radiation or micro-focus X-ray sources.

X-ray PCI was developed as a quantitative technique almost from the start, due to the use of associated phase retrieval methods [1,6,13,15]. This allowed, in particular, for development of phase-contrast computed tomography (PCT) [6,16,17]. In the context of PBI, the most successful and widespread phase retrieval technique is Paganin's method [1,18], which is based on the homogeneous variant of the Transport of Intensity equation (TIE) [13,19]. Even though this method enables recovery of the phase from the registered intensity distribution in a single transverse plane, its main strength and the key reason for its popularity is its ability to significantly increase the signal-to-noise ratio (SNR) in an image, without deteriorating the spatial resolution [18,20]. Although it may not be immediately obvious, the latter property is quite remarkable and even counter-intuitive, in view of the generic noise-resolution uncertainty (NRU) principle [21-24]. Consider an imaging setup with a certain illumination area and a fixed total number of photons, which can be determined, for example, by the object to be imaged and the allowed radiation dose. The NRU is a generalization of the simple observation that a given allocation of photons can be either distributed in a smaller number of larger detector pixels, creating high SNR but low resolution, or in a larger number of smaller detectors pixels, leading to higher resolution, but lower SNR. This trade-off has been shown to be quite fundamental: it is closely related to Shannon's information capacity in imaging and can even be used to refine the Heisenberg Uncertainty Principle (HUP) in some cases [24-26]. It was shown only relatively recently [27] that the NRU is actually violated in Paganin's method not at the "phase retrieval" stage, but rather during the forward free-space propagation of the X-ray beam from the imaged object to the detector. The NRU states that the ratio of SNR to resolution cannot exceed the total number of photons divided by the imaged area. However, this is only true with respect to statistically independent photons, and the statistical independence can increase in the sense that intensity correlations can decrease upon free-space propagation. This fact is demonstrated, for example, in the van Cittert-Zernike theorem for intensity correlations [28]. From the point of view of classical optics, it is easy to appreciate that the effective (angular) size of an incoherent X-ray source, which typically determines the degree of spatial coherence of an X-ray beam, becomes smaller with the increase of the propagation distance. The quantitative behaviour of the SNR and spatial resolution in synchrotron-based PBI experiments, and the details of the mechanism of the (beneficial) violation of the NRU in these experiments, is the main subject of the present study.

The physics of Paganin's method can be understood on the basis of general signal transmission theory. As a wavefield propagates from the exit surface of a sample to the detector entrance surface, free-space diffraction amplifies the weak high-spatial-frequency part of the signal. This PBI "encoding" step occurs before the addition of noise at the detector plane and the subsequent phase-retrieval "decoding" step. Such a process can be viewed as a particular instance of signal transmission through a noisy channel [29], for which the "noisy-channel coding theorem" [29-31] opens the logical possibility that an indirect three-step strategy—signal encoding, transmission through the noisy channel, signal decoding—may lead to SNR improvement without an increase in the total power of the signal or a decrease in its bandwidth. Letting **C** be the invertible operator that denotes the coding process, $\mathbf{N}_u$ be the operator via which uncorrelated noise is added to each pixel after the coding process, and $\mathbf{N}_c$ be the correlated noise-addition operator induced by reversing the order in which the coding and noise addition are performed, $N_c$ by definition obeys $N_u C = C N_c$, thus the similarity transformation $N_c = C^{-1} N_u C$ maps uncorrelated to correlated noise. This transformation leaves the total power of the signal and its bandwidth unchanged. However, when the phase-retrieval decoding step ($C^{-1}$) is applied to the noise in the detected image, uncorrelated (delta correlated) noise becomes correlated, decreasing the noise variance and leading to the beneficial violation of the NRU mentioned near the end of the previous paragraph.

A clarifying analogy is the Dolby noise reduction system introduced originally for compact music cassettes [32]. In Dolby noise reduction, the high temporal frequencies of an audio signal are amplified relative to their true value, before being transmitted through the noisy channel associated with recording onto magnetic audio tape, and then decoded by having the high temporal frequencies suppressed during playback via the Dolby filter. Crucially, this SNR-boosting process makes use of the prior knowledge that the high-frequency part of the target power spectrum decays more rapidly than the noise power spectrum, together with the assumption (for the purposes of our analogy) that the total power remains unchanged by the coding process. Similarly in PBI, the high-spatial-frequency components of a transmitted attenuated intensity map are amplified (in a total-power-preserving manner) relative to their true values. This amplification is enabled by the unitary-operator physics of paraxial free-space propagation from the exit surface of the sample to the entrance of the detector. The signal is then "transmitted" through the continuous-to-discrete [33] noisy channel of intensity registration via each pixel in the digital camera, and then decoded by having the high spatial frequencies suppressed during phase retrieval via the Paganin filter. This coding-transmission-decoding process results again in a suppression of noise, without a loss of the high-spatial-frequency components.

We close this introduction with a brief overview of the remainder the paper. Section 2 revises some basics of inline imaging using paraxial coherent scalar waves, including how to transition from the paraxial (wave) equation to the corresponding TIE and eikonal equations, together with a description of the form taken by the latter two equations for the case of plane waves passing through a homogeneous sample. Section 3 considers signal-to-noise ratio, spatial resolution and noise-resolution uncertainty in the context relevant to inline imaging in both two and three spatial dimensions. The beneficial violation of the NRU in propagation-based imaging is the topic of Section 4, with particular emphasis given to use of the homogeneous-sample version of the TIE in this setting. The key ideas of the preceding sections are illustrated with numerical and experimental examples, in Section 5. Section 6 contains concluding remarks.

## 2. In-line imaging

The propagation process of a scalar monochromatic electromagnetic beam in vacuum is described by the paraxial equation [34]:

$$2ik\partial_z \Psi(\mathbf{r}_\perp, z) = -\nabla_\perp^2 \Psi(\mathbf{r}_\perp, z), \qquad (1)$$

where $\Psi(\mathbf{r})\exp(ikz)$ is the complex amplitude of the beam, $\mathbf{r} = (\mathbf{r}_\perp, z)$ are Cartesian coordinates in the three-dimensional (3D) space, with $z$ being the beam propagation direction and $\mathbf{r}_\perp = (x, y)$ the position vector in transverse planes, $\nabla_\perp^2 = \partial_x^2 + \partial_y^2$ is the 2D Laplacian, $k = 2\pi/\lambda$ is the wavenumber and $\lambda$ is the wavelength. Equation (1) has the form of a 2D Schrödinger equation in free space, with $z$ in the place of the time variable, the reduced Planck constant set to one and the mass replaced by the wavenumber. The paraxial equation represents the same kind of approximation to the Helmholtz equation as the Schrödinger equation does with respect to the Klein-Gordon equation [25]. Substituting $\Psi(\mathbf{r}) = I^{1/2}(\mathbf{r})\exp[i\varphi(\mathbf{r})]$, where $I(\mathbf{r})$ is the intensity and $\varphi(\mathbf{r})$ is the phase, in the Schrödinger or, more generally, in the Klein-Gordon equation leads to the de Broglie – Bohm formalism of quantum mechanics (pilot wave theory) [35-37]. The same substitution in eq.(1) leads to the following pair of equations for the intensity and phase of a monochromatic scalar electromagnetic beam [19]:

$$k\partial_z I(\mathbf{r}_\perp, z) = -\nabla_\perp \bullet [I(\mathbf{r}_\perp, z)\nabla_\perp \varphi(\mathbf{r}_\perp, z)], \tag{2a}$$

$$2k\partial_z \varphi(\mathbf{r}_\perp, z) = -|\nabla_\perp \varphi|^2 + I^{-1/2}\nabla_\perp^2 I^{1/2}(\mathbf{r}_\perp, z), \tag{2b}$$

where $\nabla_\perp \bullet \mathbf{f} = \partial_x f_x + \partial_y f_y$ is the transverse divergence operator. Equation (2a) is called the Transport of Intensity Equation; in the pilot wave theory, it describes the propagation of particles along the field gradients. Equation (2b) is the eikonal equation which, unlike the case of ray optics in free space, has an additional "diffraction" term, $I^{-1/2}\nabla_\perp^2 I^{1/2}(\mathbf{r}_\perp, z)$. The diffraction term plays a role similar to the square of the refractive index in ray optics, both leading to bending of rays on propagation. In this sense, the diffraction term modifies the properties of space through which the rays propagate. In focal regions, the diffraction term also gives rise to the Gouy phase anomaly [38,39]. In the pilot wave theory, the diffraction term is associated with the "quantum potential", which is responsible for the reciprocal effect of particles onto the field and makes the theory non-local in nature.

A complex amplitude $\Psi(\mathbf{r}) = I^{1/2}(\mathbf{r})\exp[i\varphi(\mathbf{r})]$ is called monomorphous or homogeneous, if the proportionality coefficient $\gamma$ between the phase and the logarithm of intensity, $\varphi(\mathbf{r}) = (\gamma/2)\ln I(\mathbf{r})$, is the same at any position $\mathbf{r}$ [18]. Monomorphous amplitudes arise e.g. in the object plane after transmission of an incident plane monochromatic X-ray wave through an object consisting predominantly of a single material [18]. For such amplitudes eqs.(2a)-(2b) decouple:

$$2k\partial_z I(\mathbf{r}_\perp, z) = -\gamma \nabla_\perp^2 I(\mathbf{r}_\perp, z), \tag{3a}$$

$$2k\partial_z \varphi(\mathbf{r}_\perp, z) = (-1+\gamma^{-2})|\nabla_\perp \varphi|^2 (\mathbf{r}_\perp, z) + \gamma^{-1}\nabla_\perp^2 \varphi(\mathbf{r}_\perp, z). \tag{3b}$$

When $\gamma \gg 1$, as is typically the case in hard X-ray PBI, eq.(3b) has the form of a perturbation of the ray-optical eikonal equation (which formally corresponds to the case $\gamma = \infty$). The perturbation leads to weak bending of ray trajectories, which, in the first order of the small parameter $\gamma^{-1}$, is proportional to the curvature of the wavefront.

In the present work, we are mostly interested in the linear finite-difference approximation in eq.(3a), $\partial_z I(\mathbf{r}_\perp, z) \cong [I(\mathbf{r}_\perp, z+dz) - I(\mathbf{r}_\perp, 0)]/dz$, which corresponds to sufficiently short propagation distances $dz$ and converts eq.(3a) to the so-called homogeneous finite-difference TIE (TIE-Hom) [18]:

$$I(\mathbf{r}_\perp, R) = (1 - a^2 \nabla_\perp^2) I(\mathbf{r}_\perp, 0), \tag{4}$$

where $a^2 = \gamma R \lambda / (4\pi)$. Equation (4) provides a good approximation for the intensity distribution in in-line (propagation-based) images of monomorphous objects, if the object-plane intensity varies sufficiently slowly, so that $|a^2 \nabla_\perp^2 I(\mathbf{r}_\perp, 0)| << 1$ [18].

## 3. Signal-to-noise ratio, spatial resolution and noise-resolution uncertainty

Consider first a very simple imaging system, in which a photon fluence, $S_{in}(\mathbf{r})$, radiated from a distant source and possibly scattered by an imaged object, is incident on a position-sensitive detector. The fluence is assumed to be expressed in the number of photons per area in 2D or per volume in 3D. The former case corresponds to conventional 2D images, while the latter case may correspond, for example, to CT, where the 3D "images" can be the result of pre-processing of the initial 2D images.

It will be convenient to define the SNR and the spatial resolution via general expressions which are valid in any $n$-dimensional space. In particular, the SNR is defined as

$$SNR(\mathbf{r}) \equiv \frac{\overline{S}(\mathbf{r})}{\sigma(\mathbf{r})}, \tag{5}$$

where $\overline{S}(\mathbf{r})$ is the mean and $\sigma^2(\mathbf{r}) = \overline{[S(\mathbf{r}) - \overline{S}(\mathbf{r})]^2}$ is the variance of the fluence $S(\mathbf{r})$, with the overhead bar denoting statistical average (expectation value).

The spatial resolution can be expressed in terms of the width, defined via the second spatial integral moment, of the point-spread function (PSF), $P(\mathbf{r})$:

$$\Delta[P] \equiv \left( \frac{4\pi}{n} \frac{\int |\mathbf{r} - \overline{\mathbf{r}}|^2 P(\mathbf{r}) d\mathbf{r}}{\int P(\mathbf{r}) d\mathbf{r}} \right)^{1/2}, \tag{6}$$

where $\overline{\mathbf{r}} \equiv \int \mathbf{r} P(\mathbf{r}) d\mathbf{r}$. For non-negative functions $P(\mathbf{r})$, $\Delta[P] = [(4\pi/n) \| |\mathbf{r} - \overline{\mathbf{r}}|^2 P \|_1 / \| P \|_1]^{1/2}$, where $\| P \|_1$ is the first integral norm, corresponding to $n=1$ in the expression $\| P \|_n \equiv (\int |P(\mathbf{r})|^n d\mathbf{r})^{1/n}$. Note that in the case of symmetrical PSFs $\bar{\mathbf{r}} = 0$. In what follows, all the considered PSFs will be non-negative and normalized, such that $\|P\|_1 = 1$, unless specifically mentioned.

One popular practical approach for measuring SNR and spatial resolution is based on the assumption of "local spatial ergodicity" of the intensity distribution in an image. The latter means that, in a flat region of an image, where the intensity variation can be attributed to noise only (i.e. where the signal $\bar{S}(\mathbf{r})$ is approximately constant), a set of intensity measurements in adjacent locations can be considered as a representative sample of the statistical ensemble of intensity values at a given point of the image. In this case, the statistical mean and variance of intensity at a point $\mathbf{r}$ can be evaluated via spatial integrals over a vicinity $\Omega$ of that point:

$$\bar{S}(\mathbf{r}) = \frac{1}{|\Omega|} \int S(\mathbf{r}') d\mathbf{r}', \tag{7a}$$

$$\sigma^2(\mathbf{r}) = \frac{1}{|\Omega|} \int_\Omega [S(\mathbf{r}') - \bar{S}(\mathbf{r})]^2 d\mathbf{r}', \tag{7b}$$

where $|\Omega|$ denotes the area of $\Omega$.

Note also the following parallel between the definitions of variance of a fluence and the spatial resolution expressed by eq.(6). Let $H_\mathbf{r}(s)$ be the probability distribution function (PDF) of fluence $S(\mathbf{r})$. Then $\int s H_\mathbf{r}(s) ds = \bar{S}$ and $\int (s - \bar{S})^2 H_\mathbf{r}(s) ds = \sigma^2(\mathbf{r})$, and hence $\sigma^2(\mathbf{r}) = [1/(4\pi)] \Delta^2[H]$. In other words, the variance of a fluence is proportional to the square of the width of its PDF, which is a rather straightforward observation. The link between the variance of image intensity and the spatial resolution is exploited in the NRU principle [21-23], which is described next. The NRU states that, for any function, its spatial width and the width of its PDF cannot be made arbitrarily small at the same time [23]. This accords with the simple idea that blurring a normalised image with a unit-strength PSF will narrow the image's PDF by reducing noise, at the cost of broadening the spatial width of the image. For an intuitive pictorial representation of this key trade-off that underpins the NRU, we refer the reader to Fig. 1 in Ref. [24].

As hinted at towards the end of the previous paragraph, the detected fluence, $S_D(\mathbf{r})$, can be often expressed in the form of a convolution, $S_D = S_{in} * D$, of the fluence incident on the detector, $S_{in}(\mathbf{r})$, with the non-negative PSF of the detector, $D(\mathbf{r})$:

$$S_D(\mathbf{r}) = \int S_{in}(\mathbf{r}') D(\mathbf{r} - \mathbf{r}') d\mathbf{r}'. \tag{8}$$

We assume for now that the incident fluence is uncorrelated and is almost uniform over lengths scales comparable with the width of $D(\mathbf{r})$. In other words, we assume that the correlation length, $h$, of the incident fluence is much smaller than the width of the PSF, while $\bar{S}_{in}(\mathbf{r})$ and $\sigma^2_{in}(\mathbf{r})$ are both almost constant over distances comparable with the width of $D(\mathbf{r})$. In that case, the effect of PSF on the SNR can be described as follows [23]:

$$\frac{SNR_D(\mathbf{r})}{SNR_{in}(\mathbf{r})} = \frac{\|D\|_1}{\sqrt{h}\,\|D\|_2}. \tag{9}$$

Regarding the right-hand side of eq.(9), we note that the quantity

$$\tilde{\Delta}[P] = \left(\frac{\|P\|_1}{\|P\|_2}\right)^{2/n} \tag{10}$$

can be used as a measure of the width of the function $P(\mathbf{r})$ [23,34]. When $P$ represents the PSF of an imaging system, eq.(10) provides an alternative definition of spatial resolution, which has a different mathematical form from eq.(6), but often produces comparable results. For example, for Gaussian PSFs, $P_{Gauss}(\mathbf{r}) = (2\pi)^{-n/2} b^{-n} \exp[-|\mathbf{r}|^2/(2b^2)]$, we obtain $\Delta[P_{Gauss}] = \tilde{\Delta}[P_{Gauss}] = 2b\sqrt{\pi}$ for any $n$. In the case of a top-hat function with width $2b$, $P_{top}(\mathbf{r}) = (2b)^n \chi_{[-b,b]^n}(\mathbf{r})$, where $\chi_{[-b,b]^n}(\mathbf{r})$ is equal to 1 inside the $n$-dimensional cube $[-b,b]^n$ and is equal to 0 everywhere else, we get $\Delta[P_{top}] = 2b\sqrt{\pi/3} \cong 2.05b$ and $\tilde{\Delta}[P_{top}] = 2b$. If $P_{exp}(\mathbf{r}) = (\pi/b)\exp(-2\pi r/b)$ is an exponential distribution, then $\Delta[P_{exp}] = b\sqrt{2/\pi}$ and $\tilde{\Delta}[P_{exp}] = 2b/\pi$. Note, however, that in the case of Cauchy (Lorentzian) distributions, $P_{Cauchy}(\mathbf{r}) = (b/\pi)/(b^2 + x^2)$, we get $\Delta[P_{Cauchy}] = \infty$, while $\tilde{\Delta}[P_{Cauchy}] = 2\pi b$.

Using the definition from eq.(10), eq.(9) can be written as

$$\frac{SNR_D^2(\mathbf{r})}{\tilde{\Delta}^n[D]} = \frac{SNR_{in}^2(\mathbf{r})}{\tilde{\Delta}^n[P_{in}]}, \tag{11}$$

where $\tilde{\Delta}[P_{in}] = h$. Note that the noise correlation length, $h$, can be associated with the width of a function, $P_{in}$, whose autocorrelation is equal to the degree of spatial coherence of the incident fluence [23].

Now consider the case of linear filtering of the registered image, which can be described by the convolution $S_D * F$, i.e. by eq.(8) with the detected fluence $S_D(\mathbf{r})$ instead of the incident fluence and a non-negative filter function $F(\mathbf{r})$ instead of the PSF $D(\mathbf{r})$. According to the associativity and commutativity of the convolution operation, $S_D * F = (S_{in} * D) * F = S_{in} * (D * F) = S_{in} * (F * D)$. Therefore, if $(F*D)(\mathbf{r})$ is almost constant over distances of the order of $h$, but varies much faster than $\bar{S}_{in}(\mathbf{r})$ and $\sigma^2_{in}(\mathbf{r})$, then, arguing exactly as above, we find that after such filtering the ratio of SNR$^2$ to the effective "resolution volume" $\tilde{\Delta}_r^n[F*D]$ must remain unchanged [23]:

$$\frac{SNR_{F*D}^2(\mathbf{r})}{\tilde{\Delta}^n[F*D]} = \frac{SNR_D^2(\mathbf{r})}{\tilde{\Delta}^n[D]} = \frac{SNR_{in}^2(\mathbf{r})}{\tilde{\Delta}^n[P_{in}]} \ . \tag{12}$$

Equation (12) shows that the ratio of SNR$^2$ to the corresponding resolution volume is constant in linear shift-invariant transformations. In the case of Poisson photon statistics, this is just a restatement of the simple fact that larger effective voxels created, for example, as a result of image filtering or binning, contain more registered photons, leading to the proportionally larger SNR$^2$. Equation (12) can be alternatively understood as a statement that an increase in the photon correlation length leads to a proportional increase in the SNR, which is a well-known effect of conventional low-pass filtering of images.

The trade-off between the SNR and spatial resolution is captured in a more general context by the NRU principle [23,24] which states that, for a fixed photon fluence, any gain in the SNR is equal to or less than the corresponding increase of the minimal spatially resolvable volume. Mathematically, the NRU can be expressed as

$$\frac{\bar{S}_{in}\Delta^n[P]}{SNR_P^2} = \frac{\Delta^n[P]}{\tilde{\Delta}^n[P]} \geq C_n , \tag{13}$$

where $C_n$ is the Epanechnikov constant; $C_1 = (6/5)\sqrt{\pi/5} \cong 0.95$, $C_2 = 8/9$ and $C_3 = 60\sqrt{\pi/7^5} \cong 4/5$ [21-23]. The lower limit (equal to $C_n$) in eq.(13) is achieved for Epanechnikov PSFs, $P_{Epan}(\mathbf{r}) = A_n(1-|\mathbf{r}|^2/b^2)_+$, where $A_n$ and $b$ are constants, and the subscript "+" means that all negative values inside the brackets are replaced by zero [22]. It follows from eq.(13) that $\Delta[P] \geq C_n^{1/n} \tilde{\Delta}[P]$, i.e. $\tilde{\Delta}[P]$ generally gives a more "optimistic" estimate of the spatial resolution compared to $\Delta[P]$, which can be observed in the examples given above. Equation (13) can be extended to the cases of linear filtering of detected images, in the same way as eq.(12) extends eq.(11).

A related result is represented by the mathematical form of the classical Heisenberg Uncertainty Principle [25,41]:

$$\Delta[|U|^2]\Delta[|\hat{U}|^2] \geq 1, \tag{14}$$

where $U(\mathbf{r})$ is an arbitrary complex-valued square-integrable function and the overhead hat symbol denotes the Fourier transform, $\hat{f}(\mathbf{k}_\perp) = \iint \exp(-2\pi \mathbf{k}_\perp \cdot \mathbf{r}_\perp) f(\mathbf{r}_\perp) d\mathbf{r}_\perp$. This inequality implies that the minimal phase-space volume is bounded from below for any square-integrable function. The lower limit in the HUP is equal to one and is achieved for Gaussian functions $U(\mathbf{r}) = P_{Gauss}^{1/2}(\mathbf{r})$, for which one gets $\Delta[P_{Gauss}] = 2b\sqrt{\pi}$ and $\Delta[|P_{Gauss}^{1/2}|^2] = 1/(2b\sqrt{\pi})$. It has been shown [26] that the NRU can be used to refine the HUP, replacing the right-hand side in eq.(14) by the maximum between 1 and $C_n^{2/n} \tilde{\Delta}[|U|^2] \tilde{\Delta}[|\hat{U}|^2]$. The latter functional can be either larger or smaller than one for different functions $U(\mathbf{r})$ [26].

We are going to apply the above results to measurements of SNR and spatial resolution in 2D and 3D images. Assuming that spatial ergodicity is satisfied in a sufficiently large area of relevant images, we will use discrete analogues of eqs.(7a)-(7b), for estimation of the mean and variance of intensity in a pixel located in a flat area of the image, via the mean and variance calculated over a set of adjacent pixels. This will allow us to evaluate the SNR via eq.(5). For estimations of the spatial resolution, we will use a method based on the Fourier transform of the 2D version of eq.(8):

$$\hat{S}_D(\mathbf{k}_\perp) = \hat{S}_{in}(\mathbf{k}_\perp)\hat{D}(\mathbf{k}_\perp). \tag{15}$$

If, as assumed after eq.(8), the noisy incident fluence is uncorrelated and is almost flat within a given region of the image, then the Fourier transform of the fluence is also a flat noisy distribution. Therefore, the width of the product of the two functions in the right-hand side of eq.(15) is determined primarily by the width of the modulation transfer function (MTF), $|\hat{D}(\mathbf{k}_\perp)|$. Assuming that the PSF is Gaussian, and hence the MTF is also Gaussian, we can use the known relationship between the widths of a Gaussian distribution and its Fourier transform to evaluate the width of the PSF from the measured width of the MTF:

$$\tilde{\Delta}[D_{Gauss}] = \Delta[D_{Gauss}] = 2/\Delta[\hat{D}_{Gauss}]. \tag{16}$$

Note, however, that the relationship between the width of a function and the width of its Fourier transform, being always reciprocal in nature, does not have the same proportionality constant for all functions. In this respect, the relevant result is represented not by the HUP, eq.(14), but by the Laue inequality [42],

$$\Delta[P]\Delta[\hat{P}] \geq \Lambda_n^{1/2}. \tag{17}$$

Equation (17) can only guarantee that the product of the two widths is always larger than a certain absolute constant. The maximum possible value on the right-hand side of eq.(17) is called the Laue constant, and it is known to be in the range $0.54 < \Lambda_n < 0.85$. Unlike the case of the NRU, eq.(13), or the HUP, eq.(14), neither the exact value of $\Lambda_n$, nor the functional form of the "minimizer", corresponding to a function $D(\mathbf{r})$ for which the left-hand side of eq.(17) reaches its minimal possible value, are known. A method for measuring spatial resolution which is closely related to eqs.(15)-(17) was also developed and used previously [43-45].

## 4. Violation of NRU in propagation-based imaging

Equations (11)-(13) demonstrate that NRU is satisfied in detection and linear filtering of images. On the other hand, it is known that propagation-based (also known as in-line) imaging, which is described by eq.(4), exhibits an "unreasonable" effectiveness and can violate the NRU principle [27]. This means that PBI can produce a gain in SNR without a loss of spatial resolution or improving the spatial resolution without a loss of SNR.

Consider the case of PBI of monomorphous complex wave amplitudes, as described by TIE-Hom. Equation (4) can be trivially re-written as a convolution:

$$I(\mathbf{r}_\perp, R) = I(\mathbf{r}_\perp, 0) * T(\mathbf{r}_\perp, R), \tag{18}$$

where $T(\mathbf{r}_\perp, R) = (1 - a^2 \nabla_\perp^2) \delta(\mathbf{r}_\perp)$ and $\delta(\mathbf{r}_\perp)$ is the Dirac delta-function. It can be verified by direct calculations that the second integral moment of $T(\mathbf{r}_\perp, R)$ is equal to $-4a^2$. The fact that this second integral moment is negative means that eq.(18) acts as a deconvolution [27,46], effectively improving the spatial resolution in PBI images, $I(\mathbf{r}_\perp, R)$, in comparison with the corresponding object-plane images, $I(\mathbf{r}_\perp, 0)$.

In a real experiment, the detected intensity distribution in the object plane can often be represented as a convolution, $I(\mathbf{r}_\perp, 0) = I_{id}(\mathbf{r}_\perp, 0) * D(\mathbf{r}_\perp)$. Here $I_{id}(\mathbf{r}_\perp, 0)$ is the "ideal" object-plane intensity distribution corresponding to a delta-function detector PSF and $D(\mathbf{r}_\perp)$ is the real detector PSF. In a more general setting, with incident illumination other than a plane wave, the image blurring can also include a contribution from the spatial distribution of the source intensity [47]. After the substitution $I(\mathbf{r}_\perp, 0) = I_{id}(\mathbf{r}_\perp, 0) * D(\mathbf{r}_\perp)$, eq.(18) becomes

$$I_{id}(\mathbf{r}_\perp, R) * D(\mathbf{r}_\perp) = I_{id}(\mathbf{r}_\perp, 0) * D(\mathbf{r}_\perp) * T(\mathbf{r}_\perp, R). \tag{19}$$

While the filter function $T(\mathbf{r}_\perp, R)$ is singular, its convolution with $D(\mathbf{r}_\perp)$ can be a smooth function. For example, in the case of a 2D Gaussian PSF, $D(\mathbf{r}_\perp) = P_{Gauss}(\mathbf{r}_\perp)$, with the variance equal to $b_0^2 = 2b^2$, we obtain
$P_0(\mathbf{r}_\perp) * T(\mathbf{r}_\perp, R) = (\pi b_0)^{-2}(1 + 4a^2 b_0^{-2} - 4a^2 b_0^{-4} |\mathbf{r}_\perp|^2) \exp[-|\mathbf{r}_\perp|^2 / b_0^2]$, which is a smooth function. The second integral moment of this function is equal to $b_0^2 - 4a^2$, which can still be negative, in principle, as long as $b_0 < 2a$. However, since the second moments of the left-hand and the right-hand sides of eq.(19) must be equal, and the intensity distribution in the object plane is always non-negative, it implies that $b_0^2 - 4a^2 \geq -b_{id}^2$, where $b_{id}^2$ is the second integral moment of the function $I_{id}(\mathbf{r}_\perp, 0)$.

The fact that the filter function $T(\mathbf{r}_\perp, R)$ has negative, as well as positive values, may be presumed to be a reason why the NRU, which has been only proved for non-negative filter functions [22], does not apply to it. Let us show, however, that in fact eq.(18), and hence eq.(4), still preserve the ratio of SNR$^2$ to the spatial resolution volume. Equation (4) has an exact inverse [1,18], e.g. in the space of tempered distributions [48]:

$$I(\mathbf{r}_\perp, 0) = (1 - a^2 \nabla_\perp^2)^{-1} I(\mathbf{r}_\perp, R). \tag{20}$$

The inverse operator in the right-hand side of eq.(20) can be expressed with the help of the Fourier transform,

$$\hat{I}(\mathbf{k}_\perp, 0) = \hat{I}(\mathbf{k}_\perp, R) / (1 + 4\pi^2 a^2 k_\perp^2). \tag{21}$$

Equation (21) is known as the TIE-Hom retrieval equation or Paganin's method [18]. Due to its noise robustness and ease of practical application, it has been successfully utilized in a large variety of phase-contrast imaging scenarios using different forms of radiation and matter waves. As particular examples of this noise robustness, dose reductions by a factor of thousands or more are routinely achievable for inline imaging in computed tomography [20], thereby enabling synchrotron-based X-ray phase-contrast tomography at the rate of one thousand tomograms per second [49,50].

Taking the inverse Fourier transform of eq.(21), it is possible to re-write eq.(20) in the form of a convolution:

$$I(\mathbf{r}_\perp, 0) = I(\mathbf{r}_\perp, R) * T_{inv}(\mathbf{r}_\perp, R), \tag{22}$$

where $T_{inv}(\mathbf{r}_\perp, R) \equiv K_0(r_\perp / a)/(2\pi a^2)$ is the inverse Fourier transform of the MTF $\hat{T}_{inv}(\mathbf{k}_\perp) = 1/(1 + 4\pi^2 a^2 k_\perp^2)$ from eq.(21) and $K_0$ is the zero-order modified Bessel function of the second kind [51,52]. The second integral moment of $T_{inv}(\mathbf{r}_\perp, R)$ is equal to $4a^2$, and thus, as expected, it is equal to minus the second moment of the function $T(\mathbf{r}_\perp, R)$. As the function $K_0(\rho)$ is positive for any positive $\rho$, the convolution with $T_{inv}(\mathbf{r}_\perp, R)$ satisfies the NRU conditions. Accordingly, the transformation represented by eq.(20) increases the SNR in the exact proportion to the deterioration of the spatial resolution, so that eq.(12) holds for it (with $n = 2$). This allows us to conclude that, if eq.(4) violated the NRU, e.g. increased SNR by a factor $A \equiv SNR_{T*D} / SNR_D > \tilde{\Delta}[T*D] / \tilde{\Delta}[D]$, then by applying eq.(4) and its inverse, eq.(20), in sequence it would have been possible to increase the SNR, while leaving the intensity distribution unchanged:

$$1 = \frac{SNR_D}{SNR_D} = \frac{SNR_{T_{inv}*T*D}}{SNR_D} = \frac{SNR_{T_{inv}*T*D}}{SNR_{T*D}} \frac{SNR_{T*D}}{SNR_D} = \frac{\tilde{\Delta}[T_{inv}*T*D]}{\tilde{\Delta}[T*D]} \frac{SNR_{T*D}}{SNR_D} >$$
$$\frac{\tilde{\Delta}[T_{inv}*T*D]}{\tilde{\Delta}[T*D]} \frac{\tilde{\Delta}[T*D]}{\tilde{\Delta}[D]} = \frac{\tilde{\Delta}[T_{inv}*T*D]}{\tilde{\Delta}[D]} = \frac{\tilde{\Delta}[D]}{\tilde{\Delta}[D]} = 1. \tag{23}$$

The fourth equality in eq.(23) is the NRU applied to eq.(22), while the subsequent inequality in eq.(23) is the result of the substitution of the above assumption about the factor $A$. The obvious contradiction, $1 > 1$, obtained in eq.(23) as a result of the latter assumption, means that, in fact, the filter function $T(\mathbf{r}_\perp, R) = (1 - a^2 \nabla_\perp^2)\delta(\mathbf{r}_\perp)$ must obey the NRU, despite not being positive everywhere. The same logic can be used to prove that any filter function $F(\mathbf{r})$, whose "inverse" function, $F_{inv}(\mathbf{r})$, such that $\hat{F}_{inv}(\mathbf{k}) \equiv 1/\hat{F}(\mathbf{k})$, exists (e.g. in the space of tempered distributions), is non-negative and satisfies the conditions for "well behaved" point-spread functions described above, must also satisfy the NRU.

However, the fact that eq.(4), which is typically used to describe PBI imaging of monomorphous objects [18], satisfies the NRU, still does not actually prohibit the violation of NRU in PBI. Let us recall that eq.(4) is valid only for sufficiently slowly varying intensity distributions. In PBI experiments, the average phase function, $\varphi(\mathbf{r}) = (\gamma/2)\ln \bar{S}(\mathbf{r})$, often varies sufficiently slowly for the latter condition to be satisfied and, hence, for eq.(4) to be applicable. However, the noise component, $S(\mathbf{r}) - \bar{S}(\mathbf{r})$, typically varies very rapidly from pixel to pixel. Therefore, the free-space propagation of the photon fluence, $S(\mathbf{r})$, cannot be correctly described by eq.(4), allowing for the possibility of violation of the NRU in PBI.

An equation generalising eq.(4) to rapidly varying functions is also known [53]:

$$I(\mathbf{r}_\perp, R) = \sqrt{1+\gamma^2}\sin(\omega - \gamma^{-1}a^2\nabla_\perp^2)I(\mathbf{r}_\perp, 0), \qquad (24)$$

where $\omega \equiv \arctan \gamma^{-1}$. It is not known to us if eq.(24) has an inverse in the space of tempered distributions, which could be represented as a convolution with a positive function, as was the case with eq.(22). The Fourier transform of the convolution kernel $F(\mathbf{r}_\perp, R) = \sqrt{1+\gamma^2}\sin(\omega - \gamma^{-1}a^2\nabla_\perp^2)\delta(\mathbf{r}_\perp)$ in eq.(24) is equal to $\hat{F}(\mathbf{k}_\perp, R) = \sqrt{1+\gamma^2}\sin(\omega + 4\pi^2\gamma^{-1}a^2 k_\perp^2)$. Unlike the case of eq.(21), the latter function takes zero values at certain values of $k_\perp$. Thus, the arguments used above to prove that eq.(4) satisfies the NRU, may not apply to eq.(24). On the other hand, an inverse of a function with isolated zero values may still belong to the space of tempered distributions and may be positive, in principle [48]. Nevertheless, as demonstrated by a numerical example in the next section, for some input functions $I(\mathbf{r}_\perp, 0)$, eq.(24) amplifies noise in proportion to, or even stronger than the corresponding gain in the spatial resolution. Therefore, in such cases, eq.(24) also cannot explain the observed violation of the NRU in PBI.

A solution to the above "paradox", which suggests an explanation of the violation of NRU in PBI, can be obtained in the following way. As shown in Ref. [27], the violation of NRU in PBI can take place when the image noise is dominated by the shot noise of the photodetection process. In that case, since the paraxial free-space propagation preserves the number of photons, the SNR is the same in flat areas of the original object-plane and the propagated image-plane images. However, the (noise) variance is decreased upon the TIE-Hom retrieval, as this process corresponds to low-pass-filtering of the image fluence, according to eq.(22). For simplicity, let us consider the case where the width of the detector PSF is much smaller than the width of the TIE-Hom retrieval filter function, i.e. $\Delta^2[D] << 8\pi a^2 = 2\gamma R\lambda$. In this case, we have

$$\sigma_{ret}^2(0) = \int |T_{inv}(\mathbf{k}_\perp)|^2 W(\mathbf{k}_\perp, R)d\mathbf{k}_\perp = \int \frac{W(\mathbf{k}_\perp, 0)}{(1+4\pi^2 a^2 k_\perp^2)^2}d\mathbf{k}_\perp \cong \frac{\sigma_0^2(0)\Delta^2[D]}{\gamma R\lambda}, \qquad (25)$$

where $\sigma_{ret}^2(0)$ and $\sigma_0^2(0)$ are the variances of photon fluence in the same flat area of TIE-Hom retrieved and the original object-plane image, respectively, and $W(\mathbf{k}_\perp, z)$ denotes the power spectral density (PSD) of the image fluence in a transverse plane at distance $z$ from the object plane. In deriving eq.(25) we used the following facts.

(i) The variance of a random process is equal to the integral of its PSD [54]. In particular, $\sigma_{ret}^2(0) = \int W_{ret}(\mathbf{k}_\perp, 0) d\mathbf{k}_\perp$.

(ii) The PSD of a linearly-filtered random process is equal to the product of the original PSD and the square modulus of the Fourier transform of the filter function [54]. In particular, $W_{ret}(\mathbf{k}_\perp, 0) = |T_{inv}(\mathbf{k}_\perp)|^2 W(\mathbf{k}_\perp, R)$.

(iii) Under the paraxial conditions, the PSD in flat areas of the propagated image is the same as in the corresponding flat areas of the object-plane image, i.e. $W(\mathbf{k}_\perp, R) = W(\mathbf{k}_\perp, 0)$.

(iv) The PSD of an uncorrelated detected fluence is equal to the product of the variance and the area of the effective detector pixel [51]. As the correlation length of the detected fluence, which is equal to $\Delta[D]$, was assumed to be much smaller than the width of $T_{inv}(\mathbf{r}_\perp)$, $W(\mathbf{k}_\perp, 0)$ is almost constant over the support of $\hat{T}_{inv}(\mathbf{k}_\perp)$, and is equal to $\sigma_0^2(0)\Delta^2[D]$ [52].

(v) The integral of the function $1/(1+4\pi^2 a^2 k_\perp^2)^2$ over the 2D reciprocal space can be calculated analytically, with the result equal to $1/(4\pi a^2) = 1/(\gamma R\lambda)$.

Since the mean signal remains unchanged in flat areas of the images, eq.(25) shows that the ratio of SNR$^2$ after and before the free-space propagation followed by TIE-Hom retrieval is equal to

$$\frac{SNR_{retr}^2(0)}{SNR_0^2(0)} = \frac{\sigma_0^2(0)}{\sigma_{ret}^2(0)} = \frac{\gamma R\lambda}{\Delta^2[D]} = \frac{\gamma}{N_F}, \qquad (26)$$

where $N_F \equiv \Delta^2[D]/(R\lambda)$ is the Fresnel number corresponding to the effective pixel size in the images.

As mentioned above after eq.(19), the width of the PSF is reduced upon the free-space propagation by the same amount as it is increased upon the TIE-Hom retrieval. Therefore, after the free-space propagation followed by the TIE-Hom retrieval, the spatial resolution remains unchanged. Combining this with the increase in the SNR in accordance with eq.(26), we obtain the following expression for the gain in the SNR to resolution ratio [27,52]:

$$G_2 \equiv \left(\frac{SNR_{retr}(0)}{\tilde{\Delta}[P_{retr}]}\right) / \left(\frac{SNR(0)}{\tilde{\Delta}[D]}\right) = \sqrt{\frac{\gamma}{N_F}}, \qquad (27)$$

where $P_{retr}$ is the effective PSF after the free-space propagation followed by TIE-Hom retrieval and $D$ is the original PSF in the object plane. Note that in the derivation of eq.(25) we assumed that $N_F = \Delta^2[D]/(R\lambda) \ll 2\gamma$. Under such conditions, the gain factor can be large. The largest possible gain factor in TIE-Hom imaging was estimated to be approximately $0.3\gamma$ [52].

## 5. Numerical simulations and an experimental example

### *5.1. Numerical simulations of PBI imaging*

Here we consider the case of a plane monochromatic incident X-ray wave $\exp(ikz)$, with $k = 2\pi/\lambda$ and $\lambda = 1$ Å. The incident wave illuminated a thin homogeneous sample, with $\gamma = 100$ at the chosen wavelength. The sample was located immediately before the object plane $z = 0$. All images were assumed to be collected by an X-ray detector with a sensitive area of 10.24 cm × 10.24 cm occupied by 4,096 × 4,096 pixels. The size of the detector pixel was 25 μm × 25 μm. The X-ray transmission through the sample was modelled with the help of a function $t_{in}(\mathbf{r}_\perp)$, with the values contained in the interval (0, 0.1) and spatially distributed as in Fig.1(a). The transmitted complex amplitude in the object plane was $U_{in}(\mathbf{r}_\perp, 0) = \exp[-t_{in}(\mathbf{r}_\perp) - i\gamma t_{in}(\mathbf{r}_\perp)]$. The distribution $t_{in}(\mathbf{r}_\perp)$ represented a low-pass filtered version of imaging test patterns, which contained features with different contrasts and spatial resolutions. Low-pass filtering, using a Gaussian convolution kernel with FWHM of 4 pixels (100 μm), was applied to the original test pattern to create the function $t_{in}(\mathbf{r}_\perp)$. This was done in order to satisfy the requirement for the incident fluence to be slowly varying compared to the detector resolution, so that an adequate sampling could be ensured for subsequent simulations of free-space propagation. We inserted a square region with 1,024 × 1,024 pixels in the top right corner of the image, with $t_{in}(\mathbf{r}_\perp) = 0$ and, hence, $I(\mathbf{r}_\perp, 0) = 1$, inside this region. This flat region was used for accurate evaluation of the SNR and spatial resolution. We also inserted another square region with 1,024 × 1,024 pixels in the top left corner of the image, this region containing a pseudo-random distribution which was obtained by applying Poisson noise with the standard deviation equal to 0.1 to a uniform region of the same size, with $I(\mathbf{r}_\perp, 0) = 1$, and then low-pass filtering the result with a Lorentzian filter with the FWHM of 1054 μm. The presence of this pattern in the image allowed us to quantitatively evaluate the improvement in the spatial resolution after the free-space propagation.

In order to simulate the photon shot noise in the detected intensity, we simulated pseudo-random Poisson noise with the standard deviation equal to 20 % of the average transmitted intensity, $\exp[-2t_{in}(\mathbf{r}_\perp)]$. This corresponds to an average incident fluence of 25 photons per pixel (since $1/\sqrt{25} = 0.2$). We subsequently convolved the noisy fluence with the detector

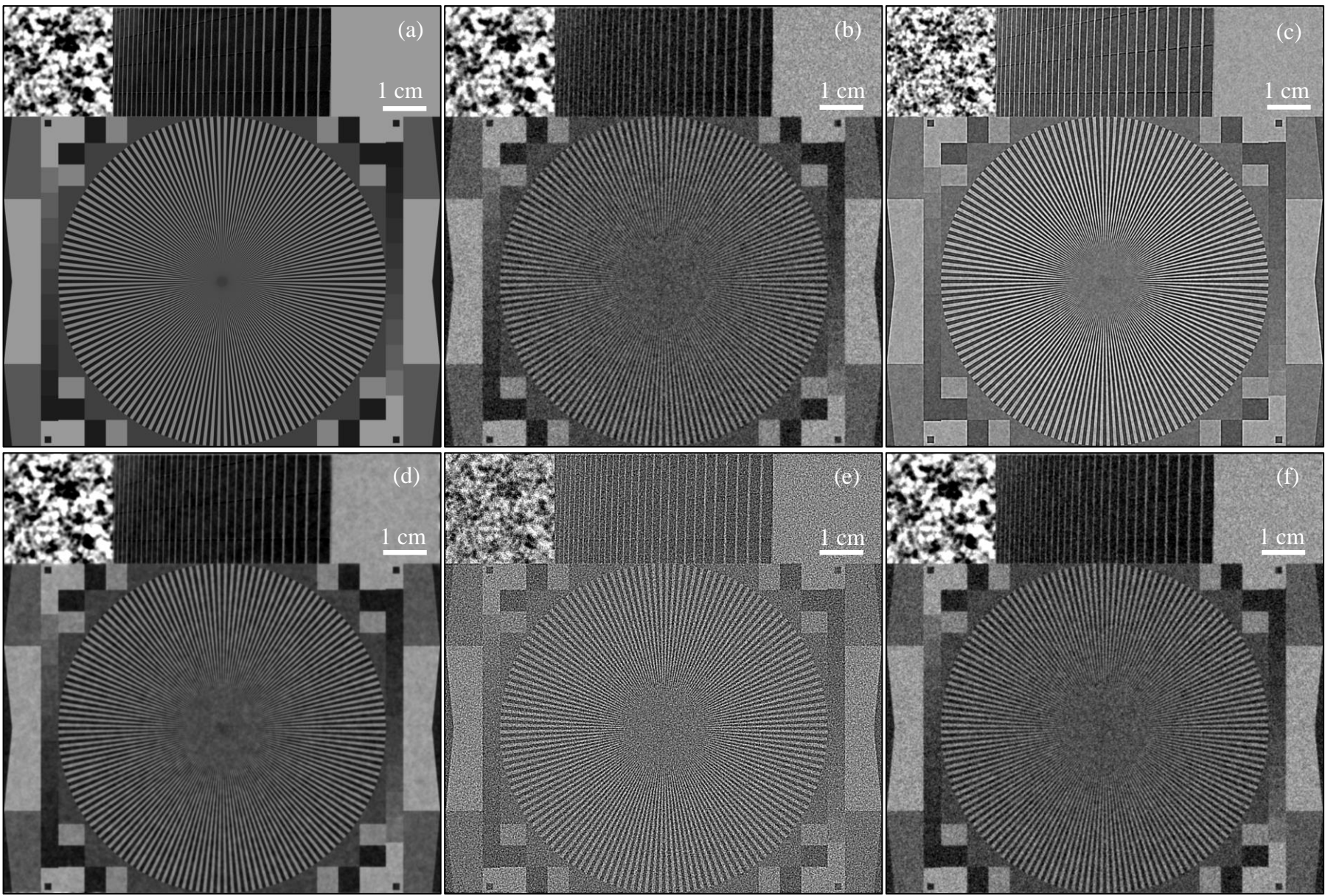

**Fig.1.** (a) Original transmission function, $-t_{in}(\mathbf{r}_\perp)$. (b) Noisy blurred detected fluence in the object plane, $S_D(\mathbf{r}_\perp,0)$. (c) Noisy blurred detected fluence in the image plane, $S_D(\mathbf{r}_\perp,R)$. (d) Distribution $S_{TIE}(\mathbf{r}_\perp,0)$, obtained by TIE-Hom phase retrieval from $S_D(\mathbf{r}_\perp,R)$. (e) Distribution $S_{D,sim}(\mathbf{r}_\perp,R)$, obtained by simulated free-space propagation of the complex amplitude produced from $S_D(\mathbf{r}_\perp,0)$. (f) Distribution $S_{TIE,sim}(\mathbf{r}_\perp,0)$, obtained by TIE-Hom retrieval from $S_{D,sim}(\mathbf{r}_\perp,R)$.

PSF, $D(\mathbf{r}_\perp)$, which was modelled as a 2D circular Gaussian distribution with $\sigma = 125$ μm. The resultant noisy blurred detected fluence, $S_D(\mathbf{r}_\perp, 0)$, is shown in Fig. 1(b). We also calculated the free-space propagation of the complex amplitude $U_{in}(\mathbf{r}_\perp)$ from the object plane $z = 0$ to the image plane $z = R = 100$ m by evaluating the corresponding Fresnel diffraction integrals. We simulated the same 20% Poisson noise in the image-plane fluence, as in the object plane, before convolving the noisy fluence with the same detector PSF as in the object plane. The resultant noisy blurred detected fluence in the image plane, $S_D(\mathbf{r}_\perp, R)$, is shown in Fig. 1(c). We then applied the TIE-Hom phase retrieval, eq.(20), to $S_D(\mathbf{r}_\perp, R)$, with the result, $S_{TIE}(\mathbf{r}_\perp, 0)$, shown in Fig. 1(d). The parameters of this simulation included a Gaussian detector PSF with $b = 125$ μm, the propagation distance $R = 10^8$ μm and the X-ray wavelength $\lambda = 10^{-4}$ μm, which corresponded to the minimal Fresnel number $N_F = 4\pi b^2 / (R\lambda) \cong 19.6$ in the in-line images. We also calculated the free-space propagation from the object plane $z = 0$ to the image plane $z = 100$ m of the complex amplitude $U_{D,sim}(\mathbf{r}_\perp, 0) = S_D^{1/2}(\mathbf{r}_\perp, 0) \exp[-i(\gamma/2) \ln S_D(\mathbf{r}_\perp, 0)]$, produced from the noisy blurred registered fluence in the detector plane. The resultant intensity distribution in the image plane, $S_{D,sim}(\mathbf{r}_\perp, R)$, can be seen in Fig. 1(e). Finally, we applied the TIE-Hom phase retrieval, eq.(20), to $S_{D,sim}(\mathbf{r}_\perp, R)$, with the result, $S_{TIE,sim}(\mathbf{r}_\perp, 0)$, shown in Fig. 1(f).

Examining Fig. 1(c), one can notice that, on a qualitative level, the forward propagation of the complex amplitude sharpened the image, without increasing noise. The image in Fig. 1(c) is substantially less noisy than that in Fig. 1(e), which contains the result of numerical free-space propagation of a complex amplitude created from the noisy detected fluence in Fig. 1(b) using the homogeneous complex amplitude $U_{D,sim}(\mathbf{r}_\perp, 0)$. As a consequence, after the application of TIE-Hom retrieval to Fig. 1(c), the result in Fig. 1(d) looks less noisy than Fig. 1(f), which contains the result of application of TIE-Hom retrieval to Fig. 1(e). The fact that Fig. 1(d) is also less noisy than the object-plane distribution in Fig. 1(b), while being as sharp as the latter, is consistent with the "unreasonable" effectiveness of PBI imaging [27]. On the other hand, the numerical free-space propagation of the complex amplitude produced from the noisy detected fluence, followed by the TIE-Hom retrieval, simply returned the noisy detected image to its original state. This is confirmed by the clear similarity of Fig. 1(f) with Fig. 1(b). The latter behaviour can be considered "reasonable", because eq.(20) is an exact inverse of eq.(4) which approximates the numerical Fresnel diffraction used to obtain Fig. 1(e) from Fig. 1(b). The same qualitative conclusions can be expressed by stating that the "true" PBI imaging (as typically implemented in experiments), consisting of free-space propagation of a complex amplitude with subsequent addition of noise and PSF blurring, followed by the TIE-Hom retrieval, violates the NRU by reducing noise without deterioration of the spatial resolution; cf. the remarks on the noisy-channel coding theorem in Section 1. At the same time, the "numerical" PBI imaging, consisting of free-space propagation of the monomorphous complex amplitude constructed from the noisy detected fluence in the object plane, followed by the TIE-Hom retrieval, conforms to the NRU, by increasing the SNR and spoiling the spatial resolution at the retrieval stage by the same amounts as the decrease in the SNR and improvement of the spatial resolution at the forward propagation simulation stage.

Therefore, these simulations are qualitatively fully consistent with the theoretical considerations presented in the previous section.

We now proceed with quantitative analysis of the SNR and spatial resolution in the images shown in Fig. 1, using software implementation of eqs.(6), (7) and (15). The following points explain our approach to this analysis and its results.

1. All measurements of the SNR have been performed in the "flat" region located in the top-right corner of the images. As expected, the SNR remained the same after the free-space propagation and it increased upon the TIE-Hom phase retrieval. The latter effect can be seen by comparing the measured values in cells c2 and d2 of Table 1, and, similarly, the values in cells e2 and f2.

**Table 1.** SNR and spatial resolution ("Res", eq.(28)) measured in images shown in Fig. 1(b)-(f) (row indices in the table correspond to the panes of Fig. 1). Spatial resolution results given in columns 3 and 5 were based on the measurement of the width of the central peak of the MTF. The results in columns 2-4 were obtained in the flat area in the top-right corner of the images, while the results in column 5 were obtained in the patterned top-left corner of the images. The results in column 6 were obtained by dividing the values in column 2 by the value in the same row of column 5.

| 1. Image | | 2. SNR | 3. Res (μm) | 4. SNR/Res (μm$^{-1}$) | 5. Res′ (μm) | 6. SNR/Res′ (μm$^{-1}$) |
|---|---|---|---|---|---|---|
| b. | $S_D(\mathbf{r}_\perp,0)$ | **91** | 257 | 0.36 | 1066 | 0.09 |
| c. | $S_D(\mathbf{r}_\perp,R)$ | 93 | 261 | 0.36 | **716** | 0.13 |
| d. | $S_{TIE}(\mathbf{r}_\perp,0)$ | **248** | 513 | 0.48 | **1068** | **0.23** |
| e. | $S_{D,sim}(\mathbf{r}_\perp,R)$ | 11 | 165 | 0.07 | 372 | 0.03 |
| f. | $S_{TIE,sim}(\mathbf{r}_\perp,0)$ | 92 | 257 | 0.36 | 1066 | 0.09 |

2. Regarding the measurements of the spatial resolution, we have found that, in practice, it is more convenient to normalize the resolution slightly differently from the normalization used in eq.(6). The following normalization leads to measured values of the spatial resolution which are close to the ones expected from a priori knowledge about the imaging conditions, particularly, in the context of experimental images considered later in the paper:

$$\text{Res}[P] = \Delta[P]/\sqrt{\pi}. \tag{28}$$

Note that eq.(28) effectively corresponds to the width of a function defined as twice the 1D standard deviation. For example, for Gaussian PSFs $P_{Gauss}(\mathbf{r}) = (2\pi)^{-n/2} b^{-n} \exp[-|\mathbf{r}|^2/(2b^2)]$ the variance is equal to $b^2$ in 1D, $2b^2$ in 2D and $3b^2$ in 3D, and in all these cases, eq.(28) gives $\text{Res}[P_{Gauss}] = 2b$. The corresponding resolution measurements in images from Fig.1 are given in Table 1, with column 3 containing the measurements performed in the top-right (flat) region and column 5 containing the measurements performed in the top-left (patterned) region. The results in column 3 reflect only the effect of the convolution with the relevant filter functions in the "flat" areas with Poisson noise. For example, the measured values of 257 µm and 261 µm in cells b3 and c3 of Table 1, respectively, agree well with the known width of the Gaussian PSF of the detector, with $2b = 250$ µm. On the other hand, the values in column 5 contain also the contribution from the "intrinsic" PSF of the pattern in the top-left corner, i.e. the Lorentzian filter with the FWHM of approximately 1054 µm. Note that the latter value is close to the measured value in cell b5 of Table 1.

3. The TIE-Hom approximation to the free-space propagation in the near-Fresnel region is described by eq.(19) with the filter function $T(\mathbf{r}_\perp, R) = (1 - a^2 \nabla_\perp^2)\delta(\mathbf{r}_\perp)$, whose second integral moment is equal to $-4a^2$. The improvement of the spatial resolution due to this filter function can be expressed as $\Delta^2[P_1] \cong \Delta^2[P_0] - 8\pi a^2$, where $P_0(\mathbf{r}_\perp)$ and $P_1(\mathbf{r}_\perp)$ are the effective PSFs in the object and in the image planes, respectively. This implies that $\text{Res}[P_1] \cong \sqrt{\text{Res}^2[P_0] - 8a^2}$. Under the conditions used in the present simulations, we get $8a^2 = (2/\pi)\gamma R \lambda \cong 636{,}620$ µm². Accordingly, the improvement in the spatial resolution in the pattern in the top left corner of the image, as a result of free-space propagation, is expected to be from $\text{Res}[P_0] = 1065$ µm (cell b5 in Table 1) approximately $\text{Res}[P_1] = \sqrt{1065^2 \, \mu m^2 - 636{,}620 \, \mu m^2} \cong 705$ µm. The latter number is close to the measured value of 716 µm in cell c5 of Table 1.

4. Similarly to the previous point, the effect of the application of TIE-Hom retrieval on spatial resolution can be estimated via the addition of the second integral moment of the corresponding filter function, $T_{inv}(\mathbf{r}_\perp, R) \equiv K_0(r_\perp/a)/(2\pi a^2)$, which is equal to $4a^2$, to the second moment of $P_1$. The resultant resolution is then equal to $\text{Res}[P_{0,retr}] = \sqrt{\text{Res}^2[P_1] + \text{Res}^2[T_{inv}]}$, where $P_1(\mathbf{r}_\perp)$ and $P_{0,retr}(\mathbf{r}_\perp)$ are the effective PSFs in the image plane and in the object plane after the TIE-Hom retrieval, respectively. Under the conditions of our simulations, we have $\text{Res}^2[T_{inv}] = 8a^2 \cong 636{,}620$ µm² and the measured value of $\text{Res}[P_1] = 716$ µm is given in cell c5 of Table 1. Hence, the expected value of $\text{Res}[P_{0,retr}]$ is $\sqrt{716^2 \, \mu m^2 + 636{,}620 \, \mu m^2} \cong 1{,}072$ µm, which is close to the measured value of 1,068 µm in cell d5 of Table 1.

5. For the parameters used in this simulation, we have $\gamma = 100$, $N_F \cong 19.6$, and hence according to eq.(27), the expected gain factor $G_2$ should be approximately $G_2 = \sqrt{100/19.6} \cong 2.26$. This theoretically predicted gain factor agrees reasonably well with the ratio of the measured values of SNR in cells b2 and d2 of Table 1: $248 / 91 \cong 2.73$, and with the measured SNR/res ratios given in cells b6 and d6: $0.23 / 0.09 \cong 2.56$. These measured gain factor values can be compared with the results obtained after the simulated free-space propagation of the complex amplitude $U_{D,sim}(\mathbf{r}_\perp, 0) = S_D^{1/2}(\mathbf{r}_\perp, 0) \exp[-i(\gamma/2) \ln S_D(\mathbf{r}_\perp, 0)]$, produced from the noisy blurred registered fluence in the detector plane, followed by the TIE-Hom retrieval. As indicated by the measured values in cells b2 and f2, as well as b6 and f6, of Table 1 we see that the gain factor in these simulations was exactly 1. This result is completely in line with the theoretical predictions given in the previous section.

*5.2. Experimental PBI imaging*

We have also measured SNR and spatial resolution in experimental X-ray images collected at the Imaging and Medical beamline (IMBL) of the Australian Synchrotron. In the experiment, a plane monochromatic X-ray beam with energy of 32 keV was used, and the propagation distances were $R = 19$ cm (approximating the object plane) and $R = 600$ cm (image plane). The detector pixel size was 100 μm, with the PSF width of approximately 160 μm. The imaged object was an excised mastectomy sample, which could be considered an approximately monomorphous sample with $\gamma = 2588$ at the specified X-ray energy [55]. Figure 2(a) shows the CT-reconstructed central slice through this sample (used here only to illustrate the actual structure of the sample), while Figs. 2 (b)-(f) contain PBI images similar to the ones in Fig.1 (b)-(f). The TIE-Hom reconstructions shown in Fig. 2 were performed with the value $\gamma = 275$, which was found to maximize the subjective radiological quality of this type of PBI reconstruction [55]. Table 2 contains the results of quantitative measurements of SNR and spatial resolution, similar to the ones given in Table 1 above, but performed on the experimental images. As the value of $\gamma = 275$ used in the TIE-Hom retrieval and in the simulated numerical free-space propagation was different from the true value of $\gamma = 2588$ for the sample, the results presented in Table 2 cannot be expected to quantitatively agree with the theoretical predictions. Nevertheless, we calculated the relevant theoretical values as a reference.

The width of the detector PSF was previously measured to be approximately 150 μm [56], the propagation distance between the two image planes in this experiment was $R = 5.81 \times 10^6$ μm and the wavelength was $\lambda = 3.875 \times 10^{-5}$ μm. Therefore, the minimal Fresnel number, corresponding to the in-line image in Fig. 2(c), was $N_F = \Delta^2 / (R\lambda) \cong 100$. The expected gain factor, $G_2 = \sqrt{\gamma / N_F}$, was approximately 1.66 for $\gamma = 275$. Comparing this with the results given in Table 2, we see that the gain in SNR evaluated as the ratio of values in cells d2 and b2 was approximately 1.62, i.e. close to the theoretically predicted value. However, comparing the corresponding spatial resolution values in cells d3 and b3, we notice that the spatial resolution was actually significantly worse in the TIE-Hom retrieved image, compared to the original "object plane" image Proj_19cm. As a result, the gain in the SNR/res ratio was actually only 0.80, as can be seen from the comparison of values in cells d4 and b4. Note, however, that this latter comparison is not very meaningful, as it is similar to the one in the numerical simulations section above, when the spatial resolution was measured in the "flat" (top right corner) areas of the images (column 3 of Table 1). As explained above, these values fail to capture the improvement in the spatial resolution upon the free-space propagation, due to the lack of image structure in such flat areas of images. The fact that some improvement in the spatial resolution was still observed in the flat areas of the experimental images, as can be seen from comparison of cells b3 and c3 of Table 2, was likely due to the effect of the X-ray source. The latter effect has not been taken into account in our theoretical analysis given above.

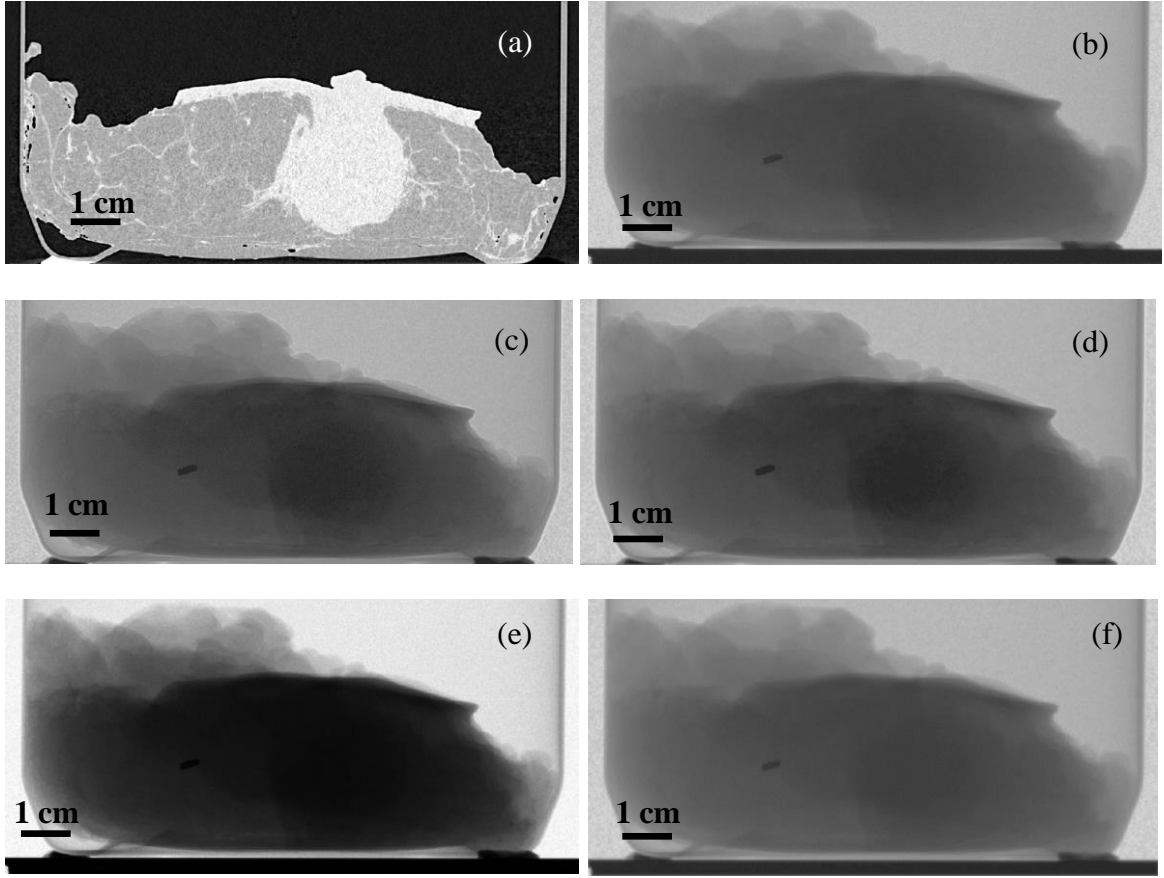

**Fig. 2.** PBI images of mastectomy sample 4704628 collected at the X-ray energy of 32 keV and the mean glandular dose of 8 mGy. (a) Reconstructed axial CT slice through the middle of the sample. (b) PBI projection at the sample-to-detector distance of 19 cm. (c) PBI projection at the sample-to-detector distance of 600 cm. (d) TIE-Hom reconstruction from (c). (e) Simulated image at the sample-to-detector distance of 600 cm obtained from the experimental image in (b). (f) TIE-Hom reconstruction from (e).

**Table 2.** SNR and spatial resolution measured in the experimental PBI images of the sample, shown in Fig. 2(b)-(f).

| 1. Image | 2. SNR | 3. Res (μm) | 4. SNR/Res (μm$^{-1}$) |
|---|---|---|---|
| b. Proj_19cm | **69** | 168 | 0.41 |
| c. Proj_6m | 60 | 138 | 0.44 |
| d. Proj_6m_TIE | **112** | 334 | 0.33 |
| e. Proj_19cm_Sim6m | 28 | 110 | 0.25 |
| f. Proj_19cm_Sim6m_TIE | 96 | 227 | 0.42 |

As the experimental images in Fig. 2 lacked suitable patterns similar to the one in the top-left corner of the images in Fig. 1, these images only allowed us to assess some qualitative effects related to the PBI theory presented above. For example, it is possible to see that sample features, such as edges and tissue interfaces in the PBI image, Fig. 2(c), do look noticeably sharper compared to the same features in the "object-plane" image in Fig. 2(b). This fact clearly indicates some improvement in the spatial resolution after the free-space propagation. Also, the reduction of noise and the commensurate improvement of SNR upon the application of TIE-Hom retrieval can be seen from the comparison of values in cell d2 vs c2 and in cell f2 vs e2. Finally, the successive application of simulated free-space propagation of the complex amplitude $U_{D,sim}(\mathbf{r}_\perp, 0) = S_D^{1/2}(\mathbf{r}_\perp, 0) \exp[-i(\gamma/2) \ln S_D(\mathbf{r}_\perp, 0)]$, produced from the noisy blurred registered fluence in the detector plane, followed by the TIE-Hom retrieval, preserved the SNR/res ratio, in agreement with the NRU. This can be seen by comparing the values in cells b4 and f4 of Table 2.

Much more detailed and comprehensive analysis of the behaviour of SNR and spatial resolution in 2D and 3D (CT) experimental PBI images will be presented in the second part of this work in a later publication.

## 6. Conclusions

It follows from previous publications (e.g. [1,2,27,52,53]) and the results presented above that the performance of Paganin's method for PBI of monomorphous objects is determined by just two key dimensionless parameters: the Fresnel number, $N_F = \Delta^2 / (R\lambda)$, and the ratio of the phase and the logarithm of intensity in the object plane, $\gamma = 2\varphi(\mathbf{r}_\perp, 0) / \ln I(\mathbf{r}_\perp, 0)$. In particular, the gain in SNR, or, equivalently, in the SNR to resolution ratio, in 2D free-space propagation followed by the TIE-Hom retrieval, is determined by the ratio of $\gamma$ and $N_F$: $G_2 = \sqrt{\gamma / N_F}$, see eqs.(26-27). Note that this gain factor depends on the dimensionality of the images: in 1D it becomes $G_1 = (\gamma / N_F)^{1/4}$ and in 3D $G_3 = \gamma / N_F$ [27,53]. The improvement of the spatial resolution upon free-space propagation is determined by the second integral moment, $-4a^2 = \gamma R \lambda / \pi = -\Delta^2 \gamma / (\pi N_F)$, of the forward TIE-Hom filter (deconvolution) function, $T(\mathbf{r}_\perp, R) = (1 - a^2 \nabla_\perp^2) \delta(\mathbf{r}_\perp)$, see eq.(19). This implies that the improvement in spatial resolution due to free-space propagation is determined by the same parameter $G_2$. More detailed and accurate theoretical estimates of the gain factor can be found in ref. [53].

The gain factor $G_n$ quantifies the "degree" of violation of NRU in PBI and, hence, the effectiveness of Paganin's method. In view of the arguments presented in Section 4 above, the latter effectiveness can be understood as the advantage that the "hardware" implementations of PBI can achieve over "software" implementations in the form of computer processing of conventional absorption-based images collected at the same radiation dose. Since the Fresnel number in PBI typically has to be much larger than unity in order to satisfy the validity conditions of the method, $\gamma$ needs to be even larger in order for the gain factor to be larger than one, i.e. for the method to work effectively. Fortunately, $\gamma = \delta / \beta$ is typically of the order of $10^3$-$10^4$ for soft biological tissues (composed of light chemical elements) when they are imaged using hard X-rays with wavelengths shorter than 1 Å, which correspond to X-ray energies higher than approximately 12 keV. Importantly, at such X-ray energies, many types of soft biological tissues can be considered approximately monomorphous. For this reason, hard X-ray PBI in general and Paganin's method in particular have become popular in biomedical applications in recent years [2-4]. The fact that the PBI gain factor can be much larger in 3D imaging [20] than in planar imaging, makes PCT a particularly attractive approach in these types of application. The method is currently being developed for medical imaging of live humans [45,55,56]. In this context, it is important to analyse the details of SNR improvement in PCT under practical conditions which require, in particular, the radiation dose and exposure time minimization. Such analysis can help researchers and engineers to optimize future medical instruments for PCT imaging using synchrotron radiation and laboratory X-ray sources. This serves as a key motivation for the detailed quantitative study presented in the second part of the present paper, which applies the theoretical framework described in the present (first) part of the paper to experimental PCT images, with particular emphasis on the role of photon-counting detectors in such applications [44,45].

**Acknowledgement**

T. E. G. and H. M. Q. acknowledge funding support of this research by the National Health and Medical Research Council, Australia (grant number APP2011204).